\documentclass[a4paper]{jpconf}
\usepackage{graphicx}
\begin{document}

\newcommand{\eli}{$\acute{{\rm E}}$liashberg }
\renewcommand{\k}{\vec{k}}
\newcommand{\kk}{\vec{k'}}
\newcommand{\q}{\vec{q}}
\newcommand{\Q}{\vec{Q}}
\renewcommand{\r}{\vec{r}}
\renewcommand{\e}{\varepsilon}
\newcommand{\ee}{\varepsilon^{'}}
\newcommand{\s}{{\mit{\it \Sigma}}}
\newcommand{\J}{\mbox{\boldmath$J$}}
\newcommand{\vv}{\mbox{\boldmath$v$}}
\newcommand{\Jh}{J_{{\rm H}}}
\newcommand{\LL}{\mbox{\boldmath$L$}}
\renewcommand{\SS}{\mbox{\boldmath$S$}}
\newcommand{\Tc}{$T_{\rm c}$ }
\newcommand{\Tcf}{$T_{\rm c}$}
\newcommand{\Hc}{$H_{\rm c2}^{\rm P}$ }
\newcommand{\Hcf}{$H_{\rm c2}^{\rm P}$}
\newcommand{\PRB}{{\it Phys. Rev.} B } 
\newcommand{\Science}{{\it Science} } 
\newcommand{\Nature}{{\it Nature} } 
\newcommand{\qf}{\vec{q}_{\rm F}}

\title{Antiferromagnetic order in the FFLO state}

\author{Youichi Yanase$^{1,2}$ and Manfred Sigrist$^2$}

\address{$^1$ Department of Physics, University of Tokyo, 
         Tokyo 113-0033, Japan}
\address{$^2$ Theoretische Physik, ETH-Honggerberg, 8093 Zurich, Switzerland}

\ead{yanase@hosi.phys.s.u-tokyo.ac.jp}

\begin{abstract}
 We investigate the antiferromagnetic (AF) order in the $d$-wave 
superconducting (SC) state at high magnetic fields. 
 A two-dimensional model with on-site repulsion $U$, inter-site 
attractive interaction $V$ and antiferromagnetic exchange interaction $J$ 
is solved using the mean field theory. 
 For finite values of $U$ and $J$, a first order transition occurs 
from the normal state to the FFLO state, while the FFLO-BCS phase transition 
is second order, consistent with the experimental results in CeCoIn$_5$. 
 Although the BCS-FFLO transition is continuous, the Ne\'el temperature of 
AF order is discontinuous at the phase boundary because the AF order 
in the FFLO state is induced by the Andreev bound state localized 
in the zeros of FFLO order parameter, while the AF order hardly occurs 
in the uniform BCS state. 
 The spatial structure of the magnetic moment is investigated for the 
commensurate AF state as well as for the incommensurate AF state. 
 The influence of the spin fluctuations is discussed for both states. 
 Since the fluctuations are enhanced in the normal state 
for incommensurate AF order, this AF order can be confined 
in the FFLO state. 
 The experimental results in CeCoIn$_5$ are discussed. 
\end{abstract}


 The FFLO superconducting state at high magnetic fields 
was predicted in 1960's by Fulde and Ferrel~\cite{rf:Fulde1964} and 
Larkin and Ovchinnikov~\cite{rf:Larkin1964}. 
 In addition to the U$(1)$-gauge symmetry the spatial symmetry 
is broken by the modulation of the SC order parameter. 
After nearly 40 years of fruitless experimental search for FFLO states
recent experiments appeared to give first evidences for such a phase
~\cite{rf:Matsudareview}.  Moreover, the FFLO phase enjoys growing interesting
in other related fields such as in cold atomic gases~\cite{rf:Zwierlein2006} 
and in high-density quark matter~\cite{rf:Casalbuoni2004}. 
 
Extensive studies of the FFLO state had been triggered by the discovery of 
a novel SC phase 
in CeCoIn$_5$~\cite{rf:Radovan2003,rf:Bianchi2003}. 
 Although several experimental results suggest the emergence of a
FFLO state here~\cite{rf:Matsudareview,rf:Watanabe2004,rf:Kakuyanagi2005,
rf:Mitrovic2006,rf:Miclea2006}, 
some NMR and neutron scattering data rather indicate the presence of 
AF order~\cite{rf:Young2007,rf:Kenzelmann2008}. 
In this paper we theoretically examine 
the possibility of the coexistence of AF order and FFLO superconductivity.

 Our theoretical analysis is based on the following model, 
\begin{eqnarray}
  \label{eq:model}
  && \hspace{-10mm}
  H=\sum_{{\k},\sigma} \varepsilon(\k) 
  c_{{\k}\sigma}^{\dag}c_{{\k}\sigma}
  + U \sum_{i} n_{{i}\uparrow} n_{{i}\downarrow} 
  + V \sum_{<i,j>} n_{i} n_{j} 
  + J \sum_{<i,j>} \vec{S}_{i} \vec{S}_{j} 
  - 2 H \sum_{i} S_{i}^{\rm z} , 
\end{eqnarray}
where $\vec{S}_{i}$ 
is the spin operator and $n_{i}$ is the number operator at site $i$. 
In order to describe the quasi-two-dimensional electronic structure 
of CeCoIn$_5$ we assume for simplicity a square lattice. 
 The bracket $<i,j>$ denotes the summation over the nearest neighbor sites. 
 The on-site repulsive interaction is given by $ U$, and
$V$ and $J$ stand for the attractive interaction and antiferromagnetic 
exchange interaction, respectively, between nearest neighbor sites. 
$V$ stabilizes the $d$-wave SC state and 
$J > 0$  takes into account 
the antiferromagnetic correlation in CeCoIn$_5$. 
 It has been experimentally shown that CeCoIn$_5$ is close to the 
quantum critical point (QCP) of AF order~\cite{rf:Matsudareview}. 
 The other candidate materials for FFLO superconductivity are also 
close to the QCP of AF order~\cite{rf:Uji2006,rf:Lortz2007,
rf:Shinagawa2007}.  
 These features, namely the $d$-wave superconductivity and AF correlation, 
can be described using the FLEX approximation for the simple 
Hubbard model~\cite{rf:yanaseFFLO}.  
 But here, we assume the interactions $V$ and $J$ to describe the FFLO 
superconductivity near the QCP within the mean field theory. 
 With the last term in eq.~(1) we include the Zeeman coupling due to the 
applied magnetic field parallel to the {\it ab}-plane. 

 We adopt the following tight-binding model, 
\begin{eqnarray}
  \label{eq:high-tc-dispersion}
  && \hspace{-10mm}
 \varepsilon(\k)=-2t(\cos k_{\rm x}+\cos k_{\rm y})
  +4t'\cos k_{\rm x} \cos k_{\rm y} -\mu, 
\end{eqnarray} 
where the unit of energy is $t=1$ and we fix $t'/t=0.25$. 
 The chemical potential enters as $\mu=\mu_{0}+\frac{1}{2} U n_{0}$ 
where $n_{0}$ is the number density for $U=V=J=0$. 
 The stable AF ordered state depends on the parameter $\mu_{0}$ 
which determines the electron number. The commensurate AF order with 
$\Q = (\pi,\pi)$ appears for $\mu_{0} = -0.8$ while an incommensurate 
AF state with $\Q = (\pi \pm \delta,\pi)$ or $\Q = (\pi, \pi \pm \delta)$
is obtained for  $\mu_{0} = -0.95$.

 We examine the model (1) within the Bogoliubov-deGennes (BdG) theory 
by taking into account the Hartree-term arising from $U$ and $J$. 
 The Hatree-term due to the attractive interaction $V$ is ignored because 
this term does not have any spin dependence as is essential for the 
following results. 
 We first determine the phase diagram for the normal, uniform BCS and 
FFLO states and later examine the magnetic instability in these states. 
  It is necessary to include both the on-site repulsion $U$ 
and antiferromagnetic $J$ in order to reproduce the experimental results 
in CeCoIn$_5$~\cite{rf:Matsudareview}. 
 For our choice of the parameters $U \sim 1$ and $J  \sim 0.6$, 
the first order phase transition occurs from the normal state to 
the FFLO state, while the BCS-FFLO transition is second order, 
consistent with the experimental results~\cite{rf:Matsudareview,
rf:Radovan2003,rf:Bianchi2003,rf:Bianchi2002}. 
 The FFLO state is suppressed for 
$T < T_{\rm c}$ if we assume $J=0$, 
while the normal-to-FFLO phase transition is 
second order around the tricritical point for $U=0$. This is
incompatible with the phase diagram of CeCoIn$_5$ indicating that 
the phase diagram of FFLO superconductivity is significantly 
affected by the electron correlation.

\begin{figure}[h]
\includegraphics[width=9cm]{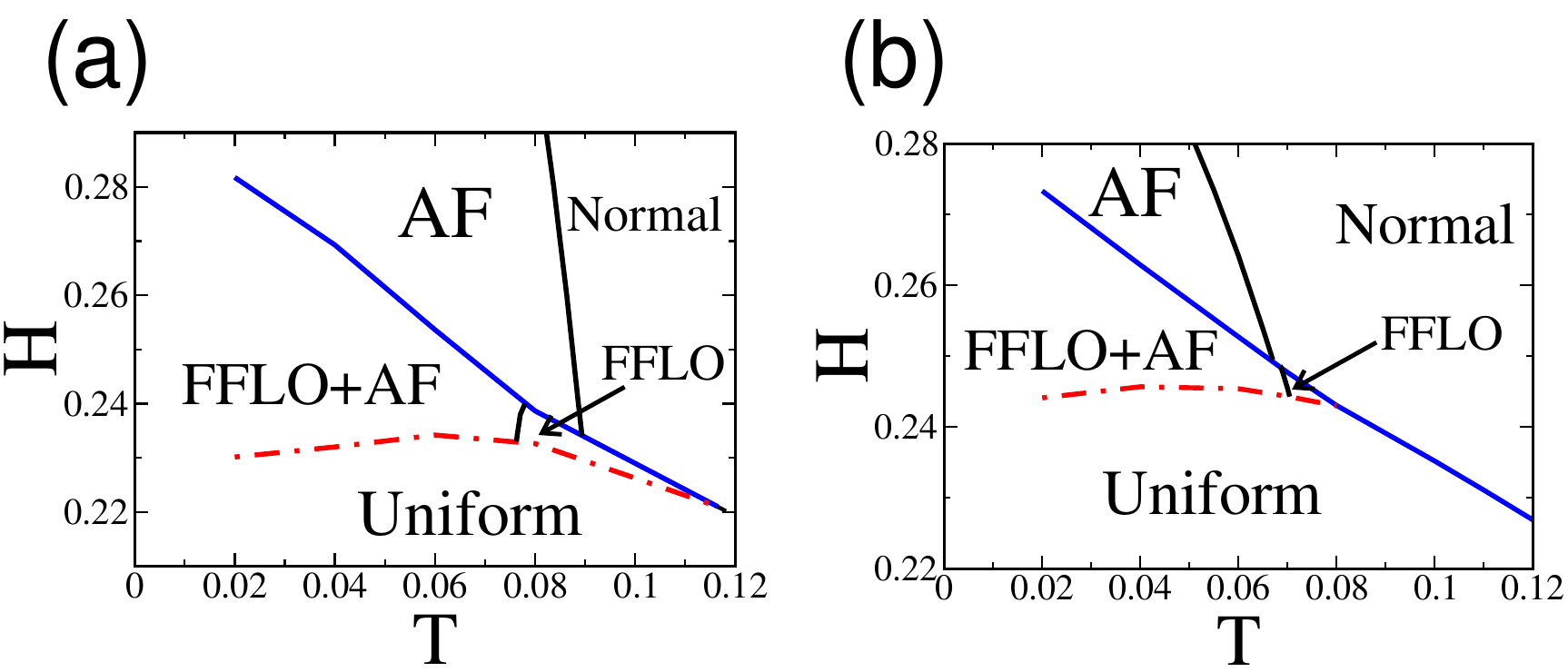}\hspace{1pc}%
\begin{minipage}[b]{6.5cm}
\caption{Phase diagram for (a) $\mu_{0} = -0.8$, $U=1$ and $J=0.6$ 
and (b) $\mu_{0} = -0.95$, $U=1.15$ and $J=0.65$. 
Blue lines show the first order phase transition to the SC state while 
red lines show the second order BCS-FFLO transition. 
Black lines show the Ne\'el temperature. 
We fix $V=-0.8$. 
}
\end{minipage}
\end{figure}

 Figures~1(a) and 1(b) show the results for $\mu_{0} = -0.8$ and 
$\mu_{0} = -0.95$, respectively. 
For both cases  AF order occurs in the FFLO and normal states, but 
is absent in the uniform BCS state. 
 The Ne\'el temperature $ T_N $ of the AF order shifts discontinuously 
at the phase boundary 
between FFLO and normal state as well as  between FFLO and uniform BCS 
state. The discontinuity at the former phase boundary is simply owing to 
the jump of SC order parameter at the first order phase transition. 
 The latter discontinuity seems more surprising because that phase boundary 
is second order. The key lies in the appearance of Andreev bound states around
the spatial nodes of the modulated SC order parameter in the FFLO state
which introduce $\pi$-phase shifts in the SC order parameter
as shown in the upper panel of Fig.~2. The continuous phase transition 
from the uniform BCS to FFLO state is associated with the nucleation 
of these domain walls like structures as shown in Fig.~2(a). 
 The Andreev bound states localized around such domain walls lead to the 
large quasiparticle density of states at the Fermi energy and induce 
the AF instability. 
 The lower panel of Fig.~2 shows that the AF staggered moment is indeed 
localized around the domain walls. 
 Because the coupling between the neighboring domain walls is weak, the 
Ne\'el temperature is almost independent of the density of walls. 
 Since the DOS is suppressed by the SC order parameter without $\pi$-phase 
shift, the AF order is absent in the uniform BCS state. 
 Therefore, the $T_N$ is discontinuous although the 
averaged magnetic moment is continuous at the phase boundary between 
the uniform BCS state and FFLO state.

\begin{figure}[h]
\vspace*{-3mm}
\includegraphics[width=16cm]{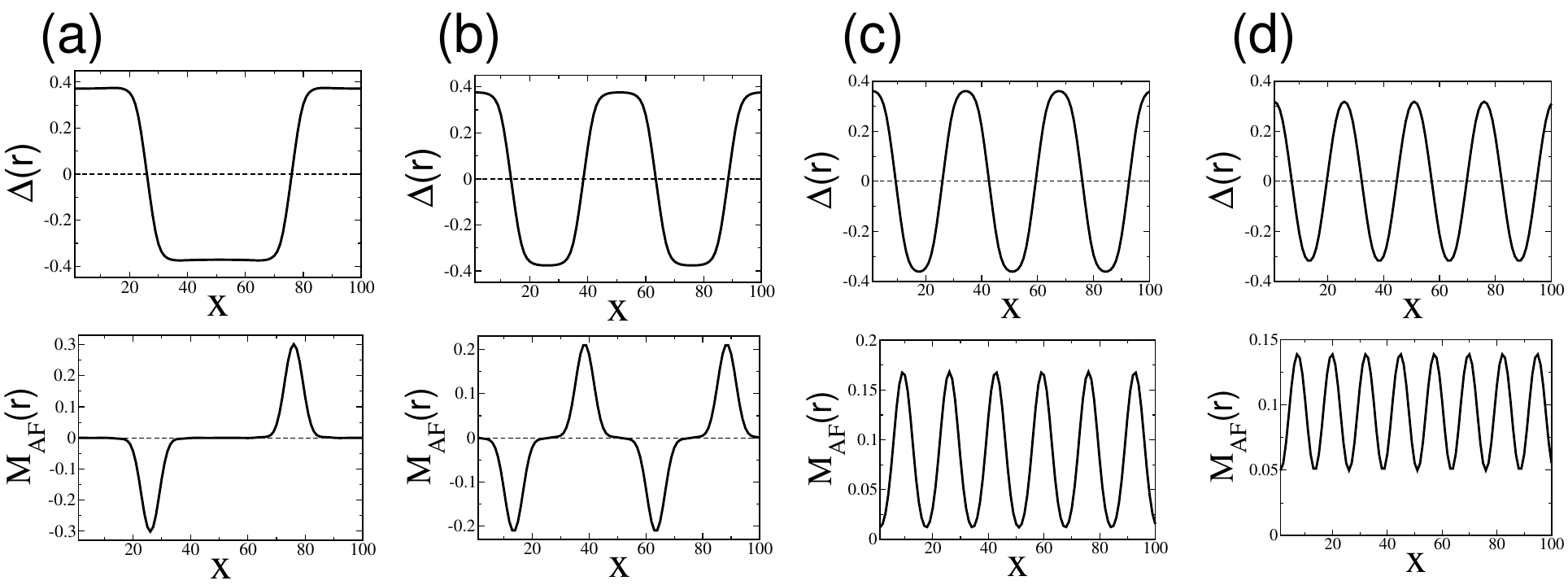}\hspace{2pc}%
\caption{Upper panel: Spatial dependences of $d$-wave SC order parameter
$\Delta(\vec{r})$ for the modulation vector of FFLO state, 
(a) $\qf$ = $(0.02 \pi,0)$, 
(b) $\qf$ = $(0.04 \pi,0)$,
(c) $\qf$ = $(0.06 \pi,0)$ and 
(d) $\qf$ = $(0.08 \pi,0)$ at the phase boundary between the FFLO and 
FFLO+AF states.  
Lower panel: Spatial dependences of AF moment 
$M_{\rm AF}(\vec{r}) = (-1)^{i+j} M(\vec{r})$ where $M(\vec{r})$ is the 
magnetic moment in the vicinity of the AF instability. 
We choose $\mu_{0}=-0.8$ so that the commensurate AF order is favored. 
 Note that $\Delta(\vec{r})$ and $M_{\rm AF}(\vec{r})$ are independent 
of $y$. The other parameters are the same as in Fig.~1(a). 
}
\end{figure}

 The discontinuity of $ T_N $ at the normal-FFLO transition 
is qualitatively different between the commensurate AF order and the 
incommensurate AF order. 
 As shown in Fig.~1, the AF order is slightly enhanced (suppressed) 
by the FFLO order in case of the incommensurate (commensurate) AF order. 
 The two dimensional spatial dependence of incommensurate AF moment is 
depicted in Fig.~3. 
 The parameters $H$ and $T$ are chosen so that the FFLO modulation vector 
is 
(a) $\qf = (0.06 \pi,0)$, 
(b) $\qf = (0.08 \pi,0)$, 
(c) $\qf = (0.031 \pi,0.031 \pi)$ and 
(d) $\qf = (0.039 \pi,0.039 \pi)$. 
 It is reasonable to assume the FFLO modulation vector parallel 
to the magnetic field $\qf \parallel \vec{H}$ 
because the vortex lattice favors this configuration~\cite{rf:Adachi2003}. 
 Therefore, Figs.~3(a,b) and 3(c,d) show the AF moment 
for the magnetic field along [100] direction, and [110] direction, 
respectively. 
 It is shown that the incommensurate structure along the nodal plane of 
FFLO order parameter is favored in both cases. 
 We examined the possibility of incommensurate AF order perpendicular 
to the nodal plane, in which the $\pi$ phase shift of AF order parameter 
is pinned by the FFLO modulation. 
 However, this AF state is less favorable than 
the AF state shown in Fig.~3 within the mean field theory.

\begin{figure}[h]
\vspace*{-6mm}
\hspace*{-20mm}
\includegraphics[width=19cm]{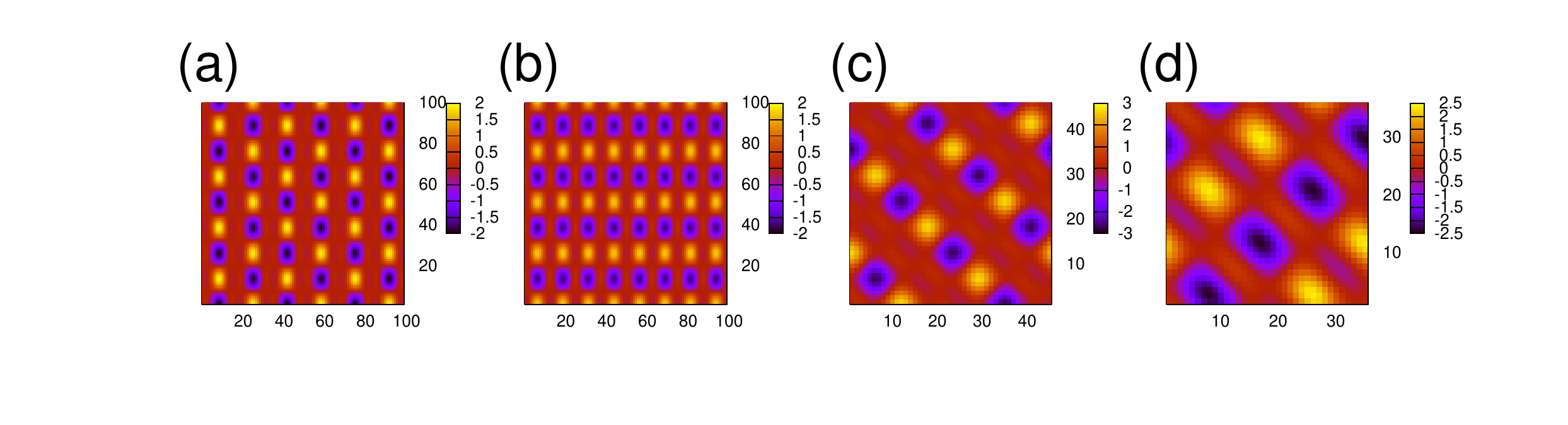}\hspace{2pc}%
\vspace*{-15mm}
\caption{Spatial dependences of AF moment 
$M_{\rm AF}(\vec{r})$ for $\mu_{0}=-0.95$ which favors the incommensurate 
AF order. The FFLO modulation vector in each figure is given in the text. 
The other parameters are the same as in Fig.~1(b). 
}
\end{figure}

 Next we discuss the experimental results for
CeCoIn$_5$~\cite{rf:Young2007,rf:Kenzelmann2008}. 
 The AF order observed at high magnetic fields and low temperatures 
might be considered as the coexistent state of FFLO superconductivity 
and AF order. 
 However, according to our result this would imply the existence of  AF order in the normal state too, 
in contrast to the experimental observation. 
While this discrepancy could be resolved using a more realistic 
electronic band structure of CeCoIn$_5$, we propose a scenario based 
on specific fluctuations of the AF wave vector $ \q $. We assume that
the spin susceptibility is enhanced at the incommensurate wave vector 
$\vec{q}=(\pi,\pi)-\vec{\delta}$ 
with a negligible anisotropy in the small $\vec{\delta}$. 
Such directional fluctuations are suppressed in the FFLO phase due to
the anisotropy introduced by the FFLO nodal planes. On the other hand,
the spin fluctuations are enhanced in the FFLO state 
because of the reduced effective dimensionality. Consequently, the realization of AF long range order
in the FFLO phase depends on a subtle balance. 
 If the directional fluctuation plays an quantitatively important 
role, the AF order in the normal state is suppressed. 
The idea of  directional fluctuation of the incommensurate AF correlation could be tested by 
neutron scattering measurements. The way  AF order and FFLO 
superconductivity coexist can be experimentally 
examined by the pressure measurements
because the AF order is likely suppressed by the pressure in analogy with the case of CeRhIn$_5$~\cite{rf:Kitaoka2005}.

\section*{Acknowledgments}
 The authors are grateful to H. Adachi, K. Izawa, M. Kenzelmann, 
Y. Matsuda, T. M. Rice, K. Yang for fruitful discussions. 
 This study has been financially supported by 
Grants-in-Aid for 
"Physics of New Quantum Phases in Superclean Materials" and 
for Young Scientists (B) from the MEXT 
and by the Center for Theoretical Studies of ETH Zurich. 
 Numerical computation in this work was carried out 
at the Yukawa Institute Computer Facility.


\section*{References}
\begin{thebibliography}{9}

\bibitem{rf:Fulde1964}
Fulde P and Ferrel R A 1964 {\it Phys. Rev.} {\bf 135} A550


\bibitem{rf:Larkin1964}
Larkin A I and Ovchinnikov Yu N 1964 {\it Zh. Eksp. Teor. Fiz.} {\bf 47} 
1136  


\bibitem{rf:Matsudareview}
Matsuda Y and Shimahara H 2007 \JPSJ {\bf 76} 051005 and 
references there in 


\bibitem{rf:Zwierlein2006}
Zwierlein M W \etal 2006
\Science {\bf 311} 492;
Partridge G B \etal 2006
\Science {\bf 311} 503  


\bibitem{rf:Casalbuoni2004}
Casalbuoni R and Nardulli G 
2004 {\it Rev. Mod. Phys.} {\bf 76} 263


\bibitem{rf:Radovan2003}
Radovan H A \etal
2003 \Nature {\bf 425} 51 

\bibitem{rf:Bianchi2003}
Bianchi A \etal
2003 \PRL {\bf 91} 187004

\bibitem{rf:Watanabe2004}
Watanabe T \etal
2004 \PRB {\bf 70} 020506(R) 


\bibitem{rf:Kakuyanagi2005}
Kakuyanagi K \etal 2005 \PRL {\bf 94} 047602;
Kumagai K \etal 2006 \PRL {\bf 97} 227002 


\bibitem{rf:Mitrovic2006}
Mitrovic V F \etal
2006 \PRL {\bf 97} 117002 


\bibitem{rf:Miclea2006}
Miclea C F \etal 2006 \PRL {\bf 96} 117001 



\bibitem{rf:Young2007}
Young B F \etal
2007 \PRL {\bf 98} 036402 


\bibitem{rf:Kenzelmann2008}
Kenzelmann M \etal 
2008 Science Express 


\bibitem{rf:Uji2006}
Uji S \etal
2006 \PRL {\bf 97} 157001 



\bibitem{rf:Lortz2007}
Lortz R \etal 
2007 \PRL {\bf 99} 187002 


\bibitem{rf:Shinagawa2007}
Shinagawa J \etal 2007 \PRL {\bf 98} 147002;  
Yonezawa S \etal 2008 Preprint arXiv:0801.0484 



\bibitem{rf:yanaseFFLO}
Yanase Y 2008 \JPSJ {\bf 77} 063705 






\bibitem{rf:Bianchi2002}
Bianchi A \etal 
2002 \PRL {\bf 89} 137002; 
Tayama T \etal 2002 \PRB {\bf 65} 180504 







\bibitem{rf:Adachi2003}
Adachi H and Ikeda R 2003 \PRB {\bf 68} 184510


\bibitem{rf:Kitaoka2005}
Kitaoka Y \etal 2005 \JPSJ {\bf 74} 186. 



\end{thebibliography}
\end{document}